\documentstyle[aps,amssymb,epsfig,multicol]{revtex}
\def\be{\begin{equation}}
\def\ee{\end{equation}}
\def\bq{\begin{eqnarray}}
\def\eq{\end{eqnarray}}
\def\bm{\begin{multicols}{2}}
\def\em{\end{multicols}}

 1

\begin{document}

\draft
\title{Finite-temperature properties of the Hubbard chain with bond-charge interaction}

\author{Fabrizio Dolcini and Arianna Montorsi}
\address{Dipartimento di Fisica and Unit\`a INFM,
Politecnico di Torino, 10129 Torino, Italy}
\date{\today}
\maketitle
\begin{abstract}
\noindent We investigate the one-dimensional Hubbard model with an
additional bond-charge interaction, recently considered in the
description of compounds that exhibit strong 1D features above the
temperature of ordered phases. The partition function of the model
is exactly calculated for a value of the bond-charge coupling; the
behavior of the specific heat and spin susceptibility as a
function of temperature is derived at arbitrary filling, and
particularly discussed across the occurring metal-insulator
transition. The results show that the bond-charge terms weaken the
spin excitations of the system.
\end{abstract}
\begin{multicols}{2}
\section{Introduction}
In recent years the discovery of materials that in some energy
regimes exhibit a strong one-dimensional character has renewed the
investigation of models of interacting electrons in low
dimensional lattices. Within this context, an increasing interest
is nowadays devoted to the effects of unconventional correlation
mechanisms, different from the usual charge-charge interaction
terms between electrons on the same site ($U$) and on neighboring
sites ($V$). In particular, models have been
considered\cite{JAKA,VOL,NIM,TINKA,AAtri}
which also account for the modification of the electron hopping
motion by the presence of particles with opposite spins
(correlated hopping); such kind of terms are also called {\it
bond-charge} interactions, since they actually describe the
interaction of charges located on bonds with those that are
located on the lattice sites.
\\The first field of application of such kind of models in condensed
matter was the description of $\pi$-electrons in conducting
polymers such as polyacetylene $\mbox{(CH)}_x$; in particular it
has been found\cite{TINKA} that, according to the strength of
bond-charge coupling the dimerization of the polymeric chain can
be enhanced or destroyed.

More recently, bond-charge models have been considered\cite{AAtri}
to explain the rich temperature-pressure phase diagram observed
for the Bechgaard salts\cite{BECH}, i.e. the linear chain organic
compounds such as $\mbox{(TMTSF)}_2 \mbox{X}$ and
$\mbox{(TMTTF)}_2 \mbox{X}$, where $\mbox{ClO}_4$ or
$\mbox{X}=\mbox{Br}$. Indeed, for these materials it has been
noticed that the spin density wave (SDW) and the superconducting
(SC) phases are adjacent, the symmetry of the SC order parameter
being of $p$-wave character, rather than $d$-wave like in
cuprates. Recent studies have suggested that the presence of bond
charge terms (whose coupling constants may depend on the
pressure) could explain the interplay between the $p$-wave~SC and
the SDW orders, with varying pressure and electron density $\rho$
(filling).
\\This idea has yielded a remarkable effort to the
investigation of electron models with correlated hopping terms,
through both analytical and numerical methods. In particular
results have been obtained for the ground state \cite{AA,SCHAD} as
well as for the low-temperature limit, by the method of
Bosonization\cite{JAKA,AAtri}, which has allowed to sketch out
the phase diagram with respect to the bond-charge coupling
constant.
\\Despite such remarkable results, a satisfactory comparison with
experimental data on Bechgaard salts has not been achieved yet.
This is mainly due to the fact that, at very low temperatures, 2D
and 3D couplings between organic chains become relevant (see for
instance the phase diagram in \cite{BECHphase}); this explains, in
pass, the occurrence of ordered phases in these compounds. As a
consequence, a one-dimensional picture for these materials is
reasonable only above some reference temperature (of the order of
$10^2$ K), which of course also depends on the pressure. In order
to compare theoretical results with experimental observations on
Bechgaard salts, it is therefore necessary to examine such models
at higher temperature, or to test whether the low-temperature
range tractable through Bosonization has a non-vanishing overlap
with the 1D region of the $P-T$ phase diagram of such materials.
\\The purpose of the present letter is to investigate the properties
for a model of bond charge interaction at finite and arbitrary
temperature. In particular, for one value of the bond-charge
coupling, we shall derive the exact behavior of thermodynamic
observables such as the specific heat and the spin
susceptibility, and discuss how the correlated hopping terms
affect the obtained shape.
\section{Model: spectrum and partition function}
The bond-charge model we shall discuss reads as follows:
\begin{eqnarray}
&\hat{\mathcal H}& = - t \sum_{\langle{i},{j}\rangle,\sigma } [1-x
\, (\hat{n}_{{i} \bar{\sigma}}+\hat{n}_{{j} \bar{\sigma}})] \,
 c_{i \sigma}^\dagger
c_{j \sigma}^{}  \, \, + \label{HAM} \\
&+& \, \, U \sum_{i=1}^L \hat{n}_{{i} \uparrow} \hat{n}_{{i}
\downarrow} -h\sum_{i=1}^L \, (\hat{n}_{{i} \uparrow}-\hat{n}_{{i}
\downarrow}) + \mu \sum_{i=1}^L \, (\hat{n}_{{i} \uparrow}
+\hat{n}_{{i} \downarrow}) \nonumber
\end{eqnarray}
In (\ref{HAM}) $c_{{i}\, \sigma}^\dagger , c_{{i} \,\sigma}^{} \,$
are fermionic creation and annihilation operators on a
one-dimensional chain of (say) $\mbox{(TMTTF)}_2 \mbox{X}$. Let
each site $i$ represent an adequately chosen unit cell\cite{CELL}
and $L$ be the total number of sites; $\sigma = \uparrow,
\downarrow$ is the spin label, $\bar{\sigma}$ denotes its
opposite, $\hat{n}^{}_{j \sigma} = c_{j \sigma}^\dagger c_{j
\sigma}^{}$ is the electron charge with spin $\sigma$, and
$\langle {i} , \, {j} \rangle$ stands for neighboring sites. At
each site $i$ four possible states are possible, which we shall
denote as follows: $|\hspace{-1mm} \uparrow
\rangle_i=c^{\dagger}_{i\uparrow}|0\rangle$, $|\hspace{-1mm}
\downarrow\rangle_i=c^{\dagger}_{i\downarrow}|0\rangle$,
$|0\rangle_i=|0\rangle$, $|\hspace{-1mm} \downarrow \uparrow
\rangle_i=c^{\dagger}_{i \downarrow} c^{\dagger}_{i \uparrow}
|0\rangle$.

The term in the first line of (\ref{HAM}) is the hopping term, and
in particular the parameter $x$ represents the bond-charge
coupling constant (for $x=0$ the ordinary Hubbard model\cite{HUB}
is recovered); the three terms in the second line respectively
describe the usual on-site Coulomb repulsion, a possible coupling
to an external magnetic field, and the chemical potential. Notice
that, similarly to \cite{VOL,AA,SCHAD} we do not consider here
the neighboring site charge-charge interaction $V$, since its
presence can be accounted for -in a first approximation- through a
renormalized value of~$U$. \\The model (\ref{HAM}) is rather
general, and is expected to capture the main effects of
bond-charge terms. In the present work we shall provide exact
results for the value:
\begin{equation}
x=1 \label{cond}
\end{equation}
For this value of the coupling constant and for zero magnetic
field ($h=0$), the exact ground phase diagram as a function of $U$
and the filling $\rho$ was obtained in \cite{AA} and \cite{SCHAD}
for open and periodic boundary conditions respectively. This
result was derived noticing that, for $x=1$, a) the term in $U$
commutes with the hopping term; b) the hopping term naturally
allows a separation of the four possible states defined above
into two groups, namely $A=\{|\hspace{-1mm}\uparrow \rangle,
|\hspace{-1mm}\downarrow \rangle \}$ and $B=\{|0 \rangle,
|\hspace{-1mm} \downarrow \uparrow \rangle\}$. In fact, for such
value of the bond-charge coupling, the hopping term actually {\it
permutes} $A$ states with $B$ states only, but not $A$ (or $B$)
states between themselves.
\\In \cite{DOMO-IJMB} these arguments have been
generalized. In the first instance, further commuting terms, other
than the on-site Coulomb repulsion, such as the magnetic field,
can be added to the hopping part. Secondly, it has been pointed
out that the properties of the hopping term of (\ref{HAM}) are
shared by a whole subclass of Extended Hubbard Models, to which
the Hamiltonian (\ref{HAM}) belongs. More explicitly, each model
within this subclass identifies a specific number and set of
Sutherland Species, i.e. the {\it groups} of states such that the
Hamiltonian only permutes the states related to different
species, leaving unaltered neighboring states that belong to the
same species; for this reason the models of this subclass have
been termed `generalized permutators'\cite{DOMO-IJMB}. The number
$n$ of Sutherland Species is by definition not greater than the
number of physical states (4 in the case of a single orbital). For
model (\ref{HAM}) the Sutherland Species are 2, and precisely $A$
and $B$.

Recognizing that a model identifies (up to some commuting terms) a
set of Sutherland Species greatly simplifies the calculation of
the partition function. The crucial point which allows that is the
use of {\it open} boundary conditions, instead of the customary
periodic ones; although in the thermodynamic limit the bulk
properties are not affected by either choice, the calculations are
more straightforward for the former. Indeed in an open
one-dimensional chain the set of eigenvalues of a generalized
permutator is equal to that of an ordinary permutator between $n$
objects, i.e. the effective dimensionality of the Hilbert space is
reduced (reduction theorem). As a consequence of that, the
degeneracy of the eigenvalues can also be computed, simply
counting the ways one can realize a given configuration of
Sutherland Species. Such observations are rather general and have
been used, for instance, to derive the exact thermodynamics of an
Extended Hubbard model of the above subclass\cite{DOMO-DM}. We
shall apply them here to obtain the partition function of the
bond-charge model~(\ref{HAM}) for the value (\ref{cond}).
\\The reduction theorem is proved when realizing that, according
to what observed above, the relative order of any sequence of
states belonging to the same species is preserved\cite{AA}; the
Hamiltonian can therefore be diagonalized within each subspace of
given set of sequences. In the case of model (\ref{HAM}), the
sequences ${\mathcal{S}}_A$ and ${\mathcal{S}}_B$ of $A$-species
and $B$-species are separately preserved. Moreover, since in this
case one has {\it two} Sutherland species, each invariant
subspace is in a one-to-one correspondence with the states of a
spinless fermion (SF) space; if the $A$ species relates to
occupied and the $B$ species to empty sites of the SF space, the
form of an effective Hamiltonian for the SF problem is that of a
tight-binding model, for the latter can be regarded to as a
permutation between occupied and empty sites. The eigenvalues
therefore read $-2 t \sum_{i=1}^{L} \cos k \, n_{k}^{A}$, with
$n_{k}^{A}$ quantum numbers valued 0 or 1, and $k=\pi n/(L+1)$,
with $n=1, \ldots, L$. The number of SF equals that of
$A$-species objects ($N_A$) and one has $\sum_{k} n_k^A=N_A$; the
number of empty sites is then $N_B=L-N_A$. Under open boundary
conditions, the specific sequence is irrelevant to the action of
the hopping term, so that all the subspaces share the same
spectrum; in general this does not hold under periodic boundary
conditions (see for instance \cite{SCHAD} or \cite{Uinf}). The
Fock space is thus reorganized in terms of states defined by
species $A$ and $B$ and their related degeneracy. \\The inclusion
of the further (commuting) terms simply lifts the degeneracy of
the eigenvalues of the hopping term, yielding a spectrum which
depends, apart from $\{ n_{k}^{A} \}$, also on $N_{\downarrow
\uparrow}$ and on $N_{\uparrow}$, the latter being the
eigenvalues of the operators $\sum_{i=1}^L \hat{n}_{{i} \uparrow}
\hat{n}_{{i} \downarrow}$ and $\sum_{i=1}^L \hat{n}_{{i}
\uparrow} (1-\hat{n}_{{i} \downarrow})$. The spectrum of model
(\ref{HAM}) thus reads
\begin{eqnarray}
E&=&E(\{n^{A}_{k}\};N_{\downarrow \uparrow};N_{\uparrow})= \label{SPE} \\
&=&\sum_{k} (\epsilon_k-\mu+h) n_k^A \,  + (U-2\mu) N_{\downarrow
\uparrow}-2h N_{\uparrow} \nonumber
\end{eqnarray}
where $\epsilon_k=-2t \cos k$; the identities
$N_\uparrow-N_\downarrow=2 N_\uparrow-N_A$ and
$N=N_\uparrow+N_\downarrow=N_A-2 N_{\downarrow \uparrow}$ have
been exploited. In the case of zero magnetic field ($h=0$) one
recovers the spectrum that was minimized in \cite{AA} at fixed
number of particles ($\mu$=const=0) to obtain the ground state
phase diagram $U$ {\it vs} $\rho$.
\\The degeneracy $g$ corresponds to the different ways one
can realize a configuration of Sutherland species, with the
constraint that the total numbers $N_{\downarrow \uparrow}$ and
$N_\uparrow$ appearing in (\ref{SPE}) remain unchanged; a simple
calculation yields
\begin{equation}\label{deg}
g(E(\{n^{A}_{k}\};N_{\downarrow \uparrow};N_{\uparrow}))={L-N_A
\choose N_{\downarrow \uparrow}} \, {N_A \choose N_{\uparrow}}
\end{equation}
The rearrangement of the Fock space deriving from the
identification of the Sutherland Species allows a straightforward
calculation of the (gran-canonical) partition function
\begin{eqnarray}
&{\mathcal Z} & = \sum_{ \{ n^{A}_k \} } \, \sum_{N_{\downarrow
\uparrow}=0}^{L-N_A} \sum_{N_\uparrow=0}^{N_A} \, g(E) e^{-\beta E
(\{n^{A}_{k}\};N_{\downarrow \uparrow};N_{\uparrow}) }\,
 = \label{zeta} \\
 & = & (1+e^{\beta(\mu-\frac{U}{2})})^{L} \prod_{k=1}^L
\left( 1+ e^{\left[- \beta \, (\epsilon_k-\mu^{*}(\mu,\beta,U,h))
\, \right]}  \right) \nonumber
\end{eqnarray} \noindent
In the second line of (\ref{zeta}) we have defined
\begin{equation}
\mu^{*}(\mu,\beta,U,h)= \mu +\frac{1}{\beta} \ln
\frac{2 \cosh
\beta h}{1+\exp{2 \beta (\mu -U/2)}}
\label{mus}
\end{equation}
$\beta=1/(k_B T)$ being the inverse temperature. Notice also that
the product over $k$ resulting in (\ref{zeta}) is in form similar
to the partition function of a tight binding model of spinless
fermions, where $\mu^{*}$ plays the role of an effective chemical
potential renormalized by the interaction $U$, the magnetic field
$h$ and the temperature itself.
\section{Results and Discussion}
By means of the partition function (\ref{zeta}) derived in the
previous section, one can calculate the thermodynamic observables
from the gran potential (per site) $\omega=-\lim_{L\rightarrow
\infty} k_B T \ln{\mathcal{Z}}$. In doing that, we have
eliminated the chemical potential $\mu$ in favor of the filling
through the relation $\rho=\partial
\omega/\partial \mu$, as usual.\\

\begin{figure}
\epsfig{file=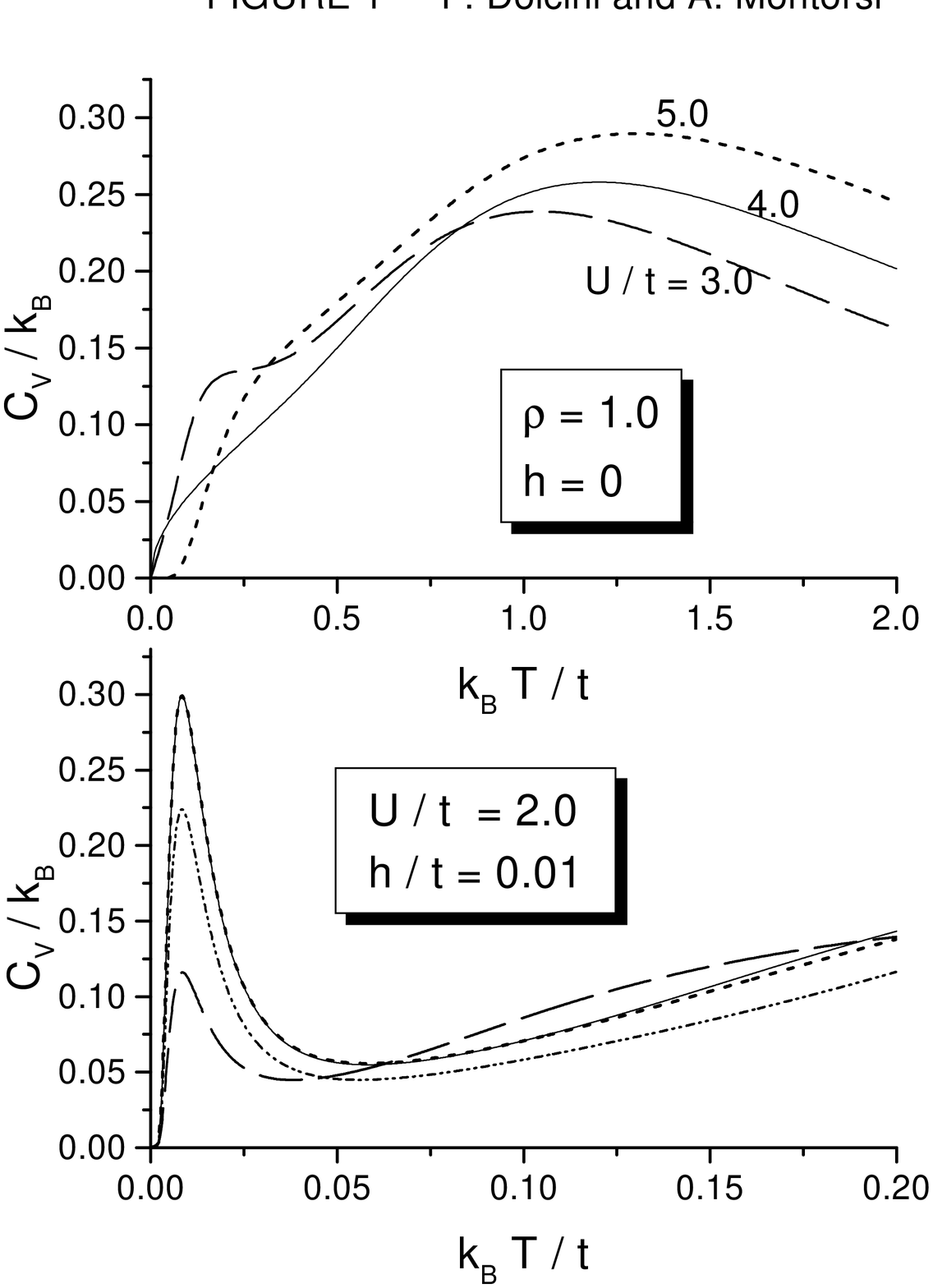,width=6.5cm,height=8.5cm,clip=}
\caption{The specific heat as a function of temperature. Top:
$\rho=1$, $h=0$: the metal-insulator transition is revealed
through the change in low-temperature behavior from linear to
exponential across the critical value $U=4t$. \\Bottom: $C_V$ for
different filling values: $\rho=0.25$ (dashed), $\rho=0.50$
(dot-dashed); $\rho=0.75$ (dotted) and $\rho=1$ (solid); a
low-temperature sharp peak emerges for non-vanishing magnetic
field.} \label{aas_fig1}
\end{figure}

In fig.\ref{aas_fig1} we have plotted the specific heat (per
site) $C_V$ as a function of the temperature. In particular, in
the top figure we have examined the case of half filling (i.e.
$\rho=1$) and zero magnetic field ($h=0$), for different values
of the on-site Coulomb repulsion $U$. One can observe that,
across the value $U/t=4$, the low-temperature behavior of $C_V$
changes from linear to exponential; explicitly, for $U<4t$ we have
\begin{equation}
C_V \sim \gamma T \hspace{0.8cm} \mbox{with} \hspace{0.8cm}
\gamma= \frac{ k_B^2 \pi}{6 t \, \sqrt{(1-(U/4t)^2)}} \, \, ,
\label{gamma1}
\end{equation}
whereas for $U>4t$
\begin{equation}
C_V \, \sim k_B \, \frac{(U-4t)^2}{8 \sqrt{\pi} t^2} (k_B
T/t)^{-3/2} \, e^{-\frac{(U-4t)}{2K_B T}} \quad. \label{exp1}
\end{equation}
This is a finite-temperature effect of a metal-insulator
transition, in accordance with the result obtained in
\cite{AA,SCHAD}, where a {\it charge} gap $\Delta_c=U-4t$ is
shown to open in the ground state for $U>4t$. We recall that for
$x=0$ (i.e. for the ordinary 1D Hubbard model) no metal-insulator
transition occurs; the bond charge term thus seems to give rise
to a finite critical value $U_c$, increasing from 0 to $4t$ as
the coupling $x$ is varied from 0 to 1. It is also worth
emphasizing that such effect is opposite to the case of higher
dimension, where the bond-charge interaction is found\cite{VOL}
to {\it lower} the critical value of the metal-insulator
transition (Gutzwiller approximation, exact in the limit
$\mbox{D} \rightarrow \infty$).
Notice that, the ratio $U/t$ is expected to scale inversely with
the pressure, since the increase of the latter roughly enhances
the hopping amplitude; as a consequence, the passage from a
metallic to an insulating state with increasing $U/t$ is in
accordance with the qualitative features of the Bechgaard salts
phase diagram\cite{BECHphase}.
\\In the bottom fig.\ref{aas_fig1}, $C_V$ is
plotted for different filling values, fixed ratio $U/t=2$ and
magnetic field $h/t=0.01$. A sharp low-temperature peak, located
at $k_B T \sim h$, is observed to emerge as soon as the magnetic
field is turned on. Interestingly, the peak becomes basically
filling independent as $\rho$ enters the range $[\bar{\rho},
2-\bar{\rho}]$, with $\bar{\rho}= \cos^{-1}(-U/4t)/ \pi$. This
amounts to the fact that, within this range of $\rho$, particles
can be added to the system only in form of singlet pairs, in
accordance with the features of the phase diagram in \cite{AA}.
\\In addition, one can show that
\begin{equation}
\lim_{T\rightarrow 0} \lim_{h\rightarrow 0} C_V/T \neq
\lim_{h\rightarrow 0} \lim_{T\rightarrow 0} C_V/T \label{non-int}
\end{equation}
differently from the ordinary Hubbard model, where the two limits
are interchangeable\cite{TAKA}. At half-filling and for
$|U+|2h||<4t$, for instance, one has $C_V \sim \gamma \, T $ with
\begin{equation}
\gamma= \frac{ k_B^2 (3 \ln^2 2+\pi^2)}{6 \pi t \,
\sqrt{(1-((U+2|h|)/4t)^2)}} \label{gamma2}
\end{equation}
Comparing eq.(\ref{gamma2}) to eq.(\ref{gamma1}), one can realize
that (\ref{non-int}) holds. Similarly, the exponential behavior,
occurring when the gap is open, is different; namely, for
$|U+|2h||>4t$
\begin{equation}
C_V \sim k_B \frac{(U+2|h|-4t)^2}{4 \sqrt{\pi} t^2} (k_B
T/t)^{-3/2} \, e^{-\frac{(U+2|h|-4t)}{2K_B T}} \quad.
\end{equation}
to be compared to eq.(\ref{exp1}).\\

\begin{figure}
\epsfig{file=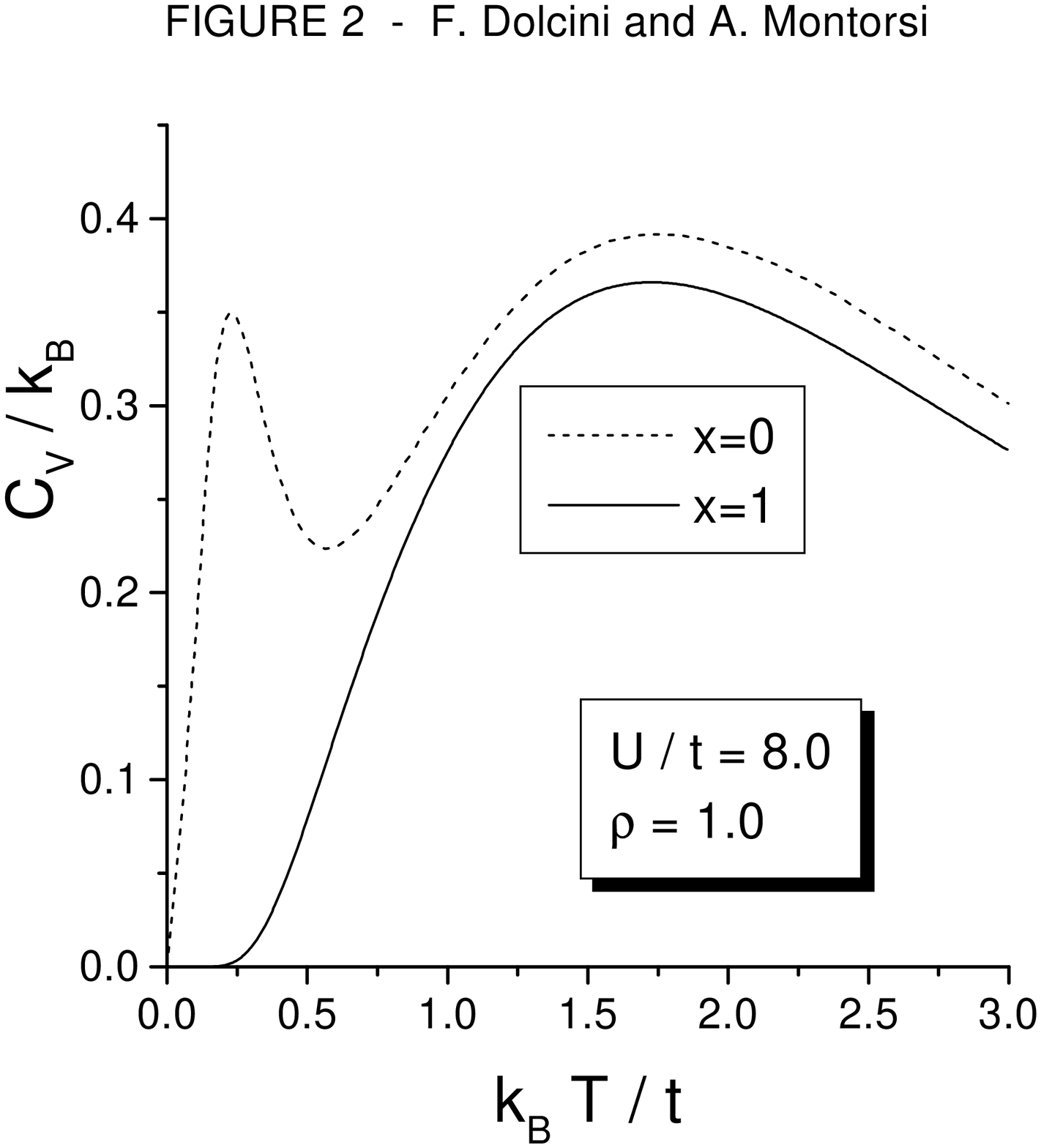,width=6.5cm,height=6.5cm,clip=}
\caption{The specific heat as a function of temperature for model
(\ref{HAM}) in the strong coupling regime ($U=8t$), at half
filling and zero magnetic field. The dotted line is the case
$x=0$ - i.e. the ordinary Hubbard model- obtained from [16], and
the solid line the case $x=1$, obtained from our exact
calculations. Continuity arguments suggest that the specific heat
for arbitrary $0 \le x \le 1$ lies between these two curves. The
low-temperature peak originating from spin excitations is
depleted by the bond-charge interaction.} \label{aas_fig2}
\end{figure}

\begin{figure}
\epsfig{file=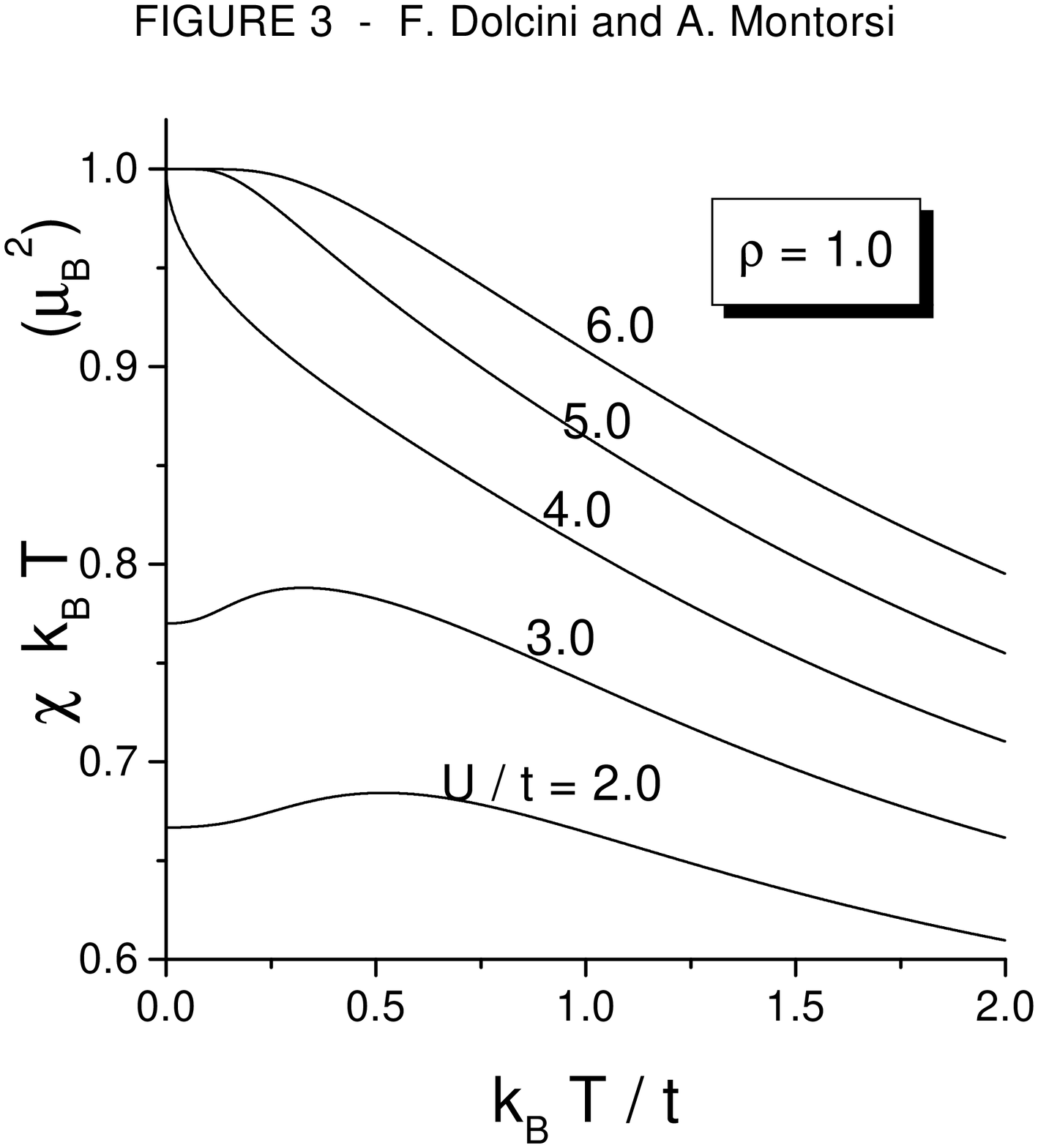,width=6.5cm,height=6.5cm,clip=}
\caption{The spin susceptibility of model (\ref{HAM}) with $x=1$
diverges like $T^{-1}$ as $T \rightarrow 0$. The quantity $\chi
k_B T$ is the integral of the spin-spin correlation function,
from fluctuation-dissipation theorem. A change in its
low-temperature behavior is observed across the metal-insulator
transition value.} \label{aas_fig3}
\end{figure}

In fig.\ref{aas_fig2} the specific heat of model (\ref{HAM}) for
$x=1$ is plotted aside the case $x=0$ (i.e. the Hubbard model) for
strong coupling, namely $U=8t$. Notice that the ordinary Hubbard
model has a low-temperature peak, whose origin (see e.g.
\cite{JKS}) is due to spin degrees of freedom; the latter being
not gapped, the low-temperature behavior of $C_V$ is linear in
spite of the fact that a charge gap is present at any
$U>0$\cite{LIWU}. In contrast, in model (\ref{HAM}), for $x=1$ the
spectrum does not carry any spin quantum number, due to the rich
symmetry of the model\cite{NOTA}; spins act therefore as a sort of
dummy variables. Although the value $x=1$ is a particular one, it
is reasonable to expect that, for continuity argument, the plot of
$C_V$ for intermediate values $0 \le x \le 1$ lies between the two
curves. As a consequence, we can infer that the effect of spin
excitations is weakened by the presence of the bond-charge
interaction, at least in the strong coupling regime. \\In order to
have a qualitative idea concerning the Bechgaard
salts\cite{BECHphase}, the temperature range of the figures
compatible with the 1D regime of e.g. $\mbox{(TMTTF)}_2 \mbox{Br}$
is $k_B T /t \gtrsim 0.3$ (indeed $t \sim 0.1$ eV and $U \sim 1$
eV).
The depletion of spin excitation is also confirmed by the behavior
of the magnetic susceptibility, defined as $\chi= \mu_B^2 \langle
\,
\partial m/ \partial h \rangle |_{h\rightarrow 0}$. The
calculation shows that $\chi$ coincides with $\mu_B^2 \rho_A/k_B
T$, where $\rho_A$ (the density of $A$-species along the chain)
is a regular function of $T$, plotted at half filling in
fig.\ref{aas_fig3}. One can observe that, differently\cite{SHIBA}
from the ordinary Hubbard model ($x=0$), the susceptibility is
divergent for $T \rightarrow 0$ with a Curie-law behavior,
reminiscent of a system of independent magnetic moments. This
also holds for any filling value, suggesting that, as $x
\rightarrow 1$, the velocity of spin excitations vanishes,
independently of $U$ and of the filling; in particular, no spin
gap is present for $x=1$ either. \\
One can now compare such results with those obtained through an a
low-energy approach. The latter is reliable only when the
interaction couplings and the thermal fluctuations are small
compared with the bandwidth $w=4t$; however, for the ordinary
Hubbard model, one can (a posteriori) extrapolate the results
concerning the formation of charge/spin gaps to the strong
coupling regime too.
\\Applying the bosonization technique to model (\ref{HAM}), one
can show (see also \cite{JAKA,AAtri}) that a) the charge sector
behaves like that of the Hubbard model, so that $U_c=0$ at
$\rho=1$, and b) the spin sector is gapless for $u'=U/t+8 x \cos(
\pi \rho/2)>0$, the spin excitations exhibiting a velocity
$v_s=v_F \sqrt{1-u'/\pi v_F}$. In particular, a linear
contribution to $C_V$ from spin excitations is thus expected for
any $x$ as long as $u'>0$. \\The exact results of the present work
(figs.\ref{aas_fig2} and \ref{aas_fig3}, and the discussion above)
indicate that any attempt to mimic the ordinary Hubbard model
extrapolating the low-energy approach to $x \rightarrow 1$ would
fail with respect to the spin channel, since the behavior is
actually quite different at $ x \simeq 1$. This also holds for the
charge channel, according to the results in \cite{AA,SCHAD} which
show that for $x=1$ the latter is gapless at half filling for
$|U| \le 4t$.\\

In conclusion, in the present paper the finite temperature
properties of the Hubbard model with bond-charge interaction have
been exactly derived for the value (\ref{cond}) of the bond-charge
coupling. The results concerning the behavior of the specific heat
and the magnetic susceptibility indicate that the bond-charge
interaction tends to suppress the spin excitations of the system.
\\Previous investigations about such model were concerned with
either the ground state or the low-temperature limit. In contrast,
our calculations are valid at any temperature. We emphasize that
for correlated quantum systems exact results are very rare at
finite temperature, even for those models that have been proved to
be integrable.

\end{multicols}

\begin{references}
\bibitem{NOTA} The degeneracy of all sequences of $A$-species
implies that the actual spin of singly occupied sites is
irrelevant.
\end{references}
\end{document}